\documentclass{actasen}

\usepackage{diagbox,multirow}
\begin{document}

\setcounter{page}{1}

\Volume{2018}{42}


\runheading{DONG Yao et al.}

\title{Tidal Evolution of Kepler lower-mass Planets$^{\dag}~ \!^{\star}$}

\footnotetext{$^{\dag}$ Supported by the
National Natural Science Foundation of China (11303102, 11273068,
11573075, 11673072), the Natural Science Foundation of Jiangsu
Province (BK20141509, BK20151607), the Foundation of Minor Planets of
the Purple Mountain Observatory.

Received 2016--09--06; revised version 2016--12--06

$^{\star}$ A translation of {\it Acta Astron. Sin.~}
Vol. 58, No. 4, pp. 31.1--31.11, 2017 \\
\hspace*{5mm}$^{\bigtriangleup}$ dongyao@pmo.ac.cn   ~~~  $^{\bigtriangleup}$$^{\bigtriangleup}$ jijh@pmo.ac.cn \\

\noindent 0275-1062/01/\$-see front matter $\copyright$ 2018 Elsevier
Science B. V. All rights reserved. 

\noindent PII: }

\enauthor{DONG Yao$^{\bigtriangleup}$$^{1,2}$\hs\hs JI Jiang-hui$^{\bigtriangleup}$$^{\bigtriangleup}$$^{1,2}$\hs\hs WANG Su$^{1,2}$}
{\up{1}Purple Mountain Observatory, Chinese Academy of Sciences, Nanjing 210008\\
\up{2}Key Laboratory of Planetary Sciences, Chinese Academy of Sciences, Nanjing 210008\\}

\abstract{The planets with a radius $<$ 4 $R$$_\oplus$ observed by the Kepler mission exhibit a unique feature, and propose a challenge for current planetary formation models. The tidal effect between a planet and its host star plays an essential role in reconfiguring the final orbits of the short-period planets. In this work, based on various initial Rayleigh distributions of the orbital elements, the final semi-major axis distributions of the planets with a radius $<$ 4 $R_\oplus$ after suffering tidal evolutions are investigated. Our simulations have qualitatively revealed some statistical properties: the semi-major axis and its peak value all increase with the increase of the initial semi-major axis and eccentricity. For the case that the initial mean semi-major axis is less than 0.1 au and the mean eccentricity is larger than 0.25, the results of numerical simulation  are approximately consistent with the observation. In addition, the effects of other parameters, such as the tidal dissipation coefficient, stellar mass and planetary mass, etc., on the final semi-major axis distribution after tidal evolution are all relatively small. Based on the simulation results, we have tried to find some clues for the formation mechanism of low-mass planets. We speculate that these low-mass planets probably form in the far place of protoplanetary disk with a moderate eccentricity via the type I migration, and it is also possible to form in situ.}


\maketitle

\section{Introduction} 

Up to now, the Kepler mission performed by NASA
has released the data of more than 4700 planetary candidates
(http://kepler.nasa.gov), most of which are considered to be real planets
\rf{1,2}. We call all the planetary candidates as planets hereafter.
About 3000 Kepler planets are distributed in the single planetary systems,
and about $\sim$ 58$\%$ of them are low-mass planets located within 0.1 au
(for example, the mass ranging from a terrestrial planet to a sub-Neptune).
Their orbital periods are about 30 d,
and the peak value of semi-major axes is distributed around 0.05 au.
The observations show that the low-mass planets are very popular in the exoplanetary systems
discovered by the Kepler telescope \rf{3-7}.
However, the discovery of the abundant \textit{Kepler} low-mass planets presents
a challenge to the current theoretical models of planet formation,
which predict a deficit of planets with a mass of 5$\sim$30 Earth mass
and an orbital period less than 50 d \rf{8}.
Therefore, the investigation of the dynamical evolution of the short-period low-mass planets
can provide a valuable clue for the study of planet formation.

There has been a lot of literature on the studies of lower-mass planets,
including the planetary composition distribution \rf{9},
mass-radius relationship \rf{10-11},
planet formation  \rf{7,12-13},
and planetary orbit characteristics\rf{5,14}.
Interestingly, all these studies are related to the formation mechanism of lower-mass planets.
Furthermore, the studies of planet occurrence predict that the
planetary occurrence rate rises with the decrease of planetary mass\rf{8,14}.
In this sense, the higher planetary occurrence rate implies some important information of planet formation.

Currently, the generally accepted formation mechanisms of
these short-period low-mass planets
include the migration in the far place of protoplanetary disk\rf{15},
planet-planet scattering \rf{16}
and \textit{in situ} formation \rf{17}.
The tidal effect plays an essential role in reshaping the
final orbital configuration of a planet
in the late stage of planet formation\rf{18-19}.
The planetary tide caused by the host star can make the planetary orbit decay and circularized,
at the same time, the stellar tide caused by the planet makes the planet continuously migrate inwards
until dropping in the Roche limit.
Most of previous studies mainly focused on the tidal evolution of a particular
single or multiple planetary system\rf{20-25}, and gave the process of planetary orbit decay and circularization.
In this work, we make the numerical simulation of tidal evolution on a large number of
short-period low-mass planets, based on the
various initial distributions of $a$ and $e$,
to find out the finally evolved distribution of orbital semi-major axes,
and attempt to reveal the possible formation history of this kind of planets.
According to the obtained results,
we speculate that this kind of planets cannot form directly at the places
very close to their host stars (e.g. within 0.05 au), but possibly pile-up at the places near 0.1 au at first
before they drop to the close orbits under the tidal action.

\section{METHOD}

\subsection{Dynamical Model}

We adopt the model of tidal evolution that the mean variation rates of
orbital elements caused by the tidal effect in a single-planet system are given by

\begin{equation}
 < \frac{da}{dt} > = -\frac{4}{3}na^{-4}\hat{s}[(1+23e^2)+7e^2D]
 \end{equation}
 and
 \begin{equation}
 < \frac{de}{dt} > = -\frac{2}{3}nea^{-5}\hat{s}[9+7D]
 \end{equation}
 where $D\equiv$ $\hat{p}/(2\hat{s}$), and
 \begin{equation}
\hat{s} \equiv \frac{9}{4}\frac{k_*}{Q_*}\frac{m_p}{m_*}R_*^5, \quad\hat{p} \equiv \frac{9}{2}\frac{k_p}{Q_p}\frac{m_*}{m_p}R_p^5
 \end{equation}
represent the effects of stellar and planetary tides, respectively \rf{22}.
In the above equations, $m$, $R$ are mass and radius,
where the subscripts $*$ and $p$ denote the stellar and planetary parameters,
$k$ and $Q$ are the Love number and tidal dissipation parameter,
$a$, $e$ and $n$ are the semi-major axis, eccentricity and mean orbital motion, respectively.
In general, the stellar tide
is much weaker than that of the planet, because that the mass ratio of the
planet and star is small, and $Q_*$ is much larger than $Q_p$ for lower-mass planets.
In the study of tidal evolution of a planetary system, we generally use the modified tidal dissipation coefficient
$\textbf{$Q^{\prime}$ $\equiv$ 3$Q$/2$k$}$
coupling with the Love number\rf {19}.
The above equations are valid
when the star has a rotation velocity much slower than the planet's orbital motion,
and based on the implicit hypothesis that the tidal dissipation coefficient is a
constant during the evolution.

\subsection{Initial Value Setting}

We carry out the numerical simulations of a large sample of terrestrial planets closely orbiting
around solar-like stars, using the seventh-order Runge
Kutta integrator. Based on the sample of planets  with a radius less
than 4 $R_\oplus$ and a semi-major axis within 0.3 au in the Kepler single planetary systems,
we have assumed a large number of initial planetary systems.
The planetary radius is assumed to satisfy a Rayleigh distribution\rf{26},
which is characterized by the Rayleigh
coefficient $\sigma_R $ = 1.3 $R_\oplus$,
or by the mean radius $\bar{R}$ = $\sigma_R \sqrt{\pi/2}$.
The planetary mass cannot be determined by the transit method,
but can be obtained by fitting the mass-radius relation of
$M_p/M_\oplus$ = 2.69 $\times$ ($R_p/R_\oplus$)$^{0.93}$
\rf{11}. As a result, the planetary mass is below 20 $M_\oplus$.
Herein, the star is assumed to be FGK type, and the stellar mass and radius are assumed to be 1 $M_\odot$ and 1 $R_\odot$, respectively.
We use $Q^{\prime}_p$ $=$ 10 to save the calculation time,
because that this parameter has no effect on the final orbit location of a planet in the secular dynamical evolution\rf{23-24}.
The parameter $Q^{\prime}_*$ is crucial to reform the final planetary configuration,
but it is limited by the observations.
Herein we adopt a generally accepted value of $Q^{\prime}_*$ = 10$^7$ \rf{27}.
The initial semi-major axis and orbital eccentricity distributions are
assumed to be the Rayleigh distributions \rf{26},
with the Rayleigh coefficients $\sigma_a$ $=$ 0.03, 0.05, 0.08 (Fig.1)
and $\sigma_e$ $=$ 0.05, 0.1, 0.2, 0.3, 0.4, respectively.
Therefore, we have three semi-major axis regions from several stellar radius to 0.11 au, 0.18 au and 0.29 au
, of which $\sim$ 90$\%$ of planets are piled-up in the peaks.
Our numerical simulations show that
most planetary orbits will be circularized in several million years but continue to reduce within the solar life under the action of stellar tide.
When the evolution times are randomly selected in the range of 1$-$6 Gyr, most planets will experience the orbit circularization in the process of stellar tidal evolution, whereas a small part of planets will fell in the Roche limit
$a_{R}$ = 2.16 $R_p(M_*/M_p)^{1/3}$, and be disintegrated by the tide\rf{28}.
In total, we have performed 15 groups of numerical simulations, each group
 contains 2800 planetary systems, and finally about 10$\%$
is disintegrated by the tide. At the same time, we compare the simulated results with
the Kepler observation sample of 1939 \textit{Kepler} planets (within 0.3 au) with a radius of $R_p$ $<$ 4 $R_\oplus$
in single planetary systems.

\begin{figure}[tbph]
\centering
\includegraphics[scale=1.2]{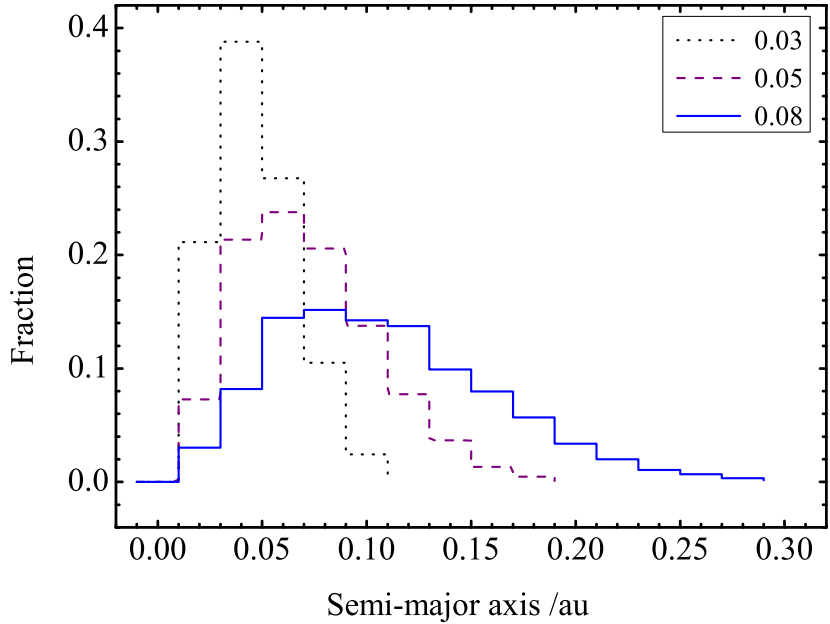}
\caption{The initial semi-major axis distributions with Rayleigh coefficients
  $\sigma_a$ of 0.03 (dotted line), 0.05 (dashed line) and 0.08 (solid line) in the simulations.}
    \end{figure}

\section{RESULTS}

It is predicted by the tidal theory that the evolution of planetary orbital eccentricity will
affect the final orbital semi-major axis, hence we make first the numerical simulation for the tidal evolutions of planet's orbital semi-major axis and eccentricity (Fig.2).
The result shows that in the solar-like planetary system,
the circularization timescale of an Earth-like planet is about 30 Myr (the upper panel of Fig.2),
simultaneously, the planet's semi-major axis decreases with the orbital circularization
due to the action of the planetary tide (the lower panel in Fig.2).
In the following we will present the results of evolution for the different initial orbits.

The final semi-major axis distributions
in the case of $\sigma_a$ = 0.03 are shown in Fig.3,
under the initial eccentricity distributions of
$\sigma_e$ $=$ 0.05, 0.1, 0.2, 0.3 and 0.4, respectively.
The results show that all the final semi-major axis distributions are peaked at about 0.03 au,
exceeding the initial peak value by a fraction of 29$\%$,
because most of planets beyond 0.03 au have been piled up
near the peak locations after a secular tidal evolution.
Furthermore, more planets will migrate to the peak locations with an greater initial eccentricity.
It seems that the final distributions have no much differences when the initial eccentricity is below 0.1.
The results can be well explained by the classical tidal theory
that tidal effect plays a very important part in
the dynamical evolution for the planets with the orbits within 0.05 au \rf{29}.
However, the simulation results are not in good agreement with the semi-major axis distribution given by
the observation. When the eccentricity is in the range from 0.05 to 0.3,
about 9$\%$-29$\%$ of planets are migrated within 0.03 au by tidal forces.
Simultaneously, more than 15$\%$ of planets enter into the Roche limit to be destroyed by tides.
Hence, we infer that most short-period Earth-like planets are impossible
to form \emph{in situ} directly, but may form in and migrate from a far place of protoplanetary disk.
\begin{figure}[tbph]
\centering
\includegraphics[scale=0.50]{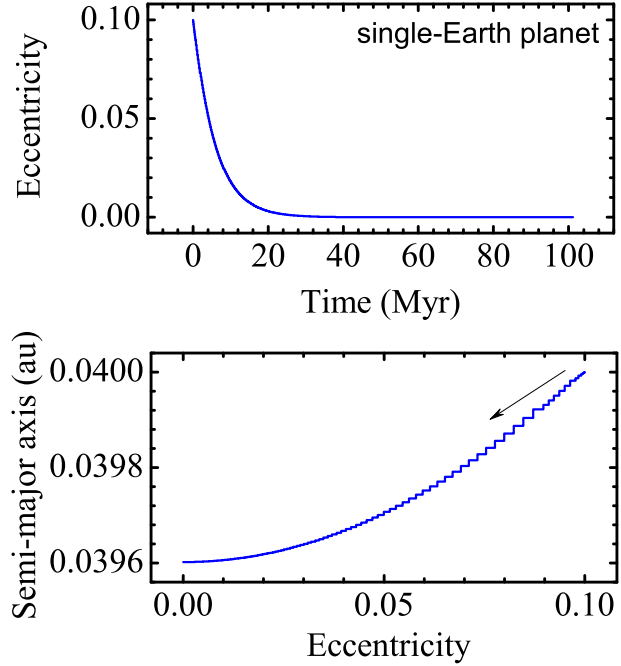}
\vspace{-2mm}
\caption{Tidal evolution of the semi-major axis and eccentricity in a single-Earth planetary system.
  The upper panel shows the evolution of the orbital eccentricity,
  and the lower panel shows the semi-major axis varying with eccentricity (the arrow indicates the evolution direction).}
    \end{figure}

\begin{figure}[tbph]
\centering
\includegraphics[scale=1.2]{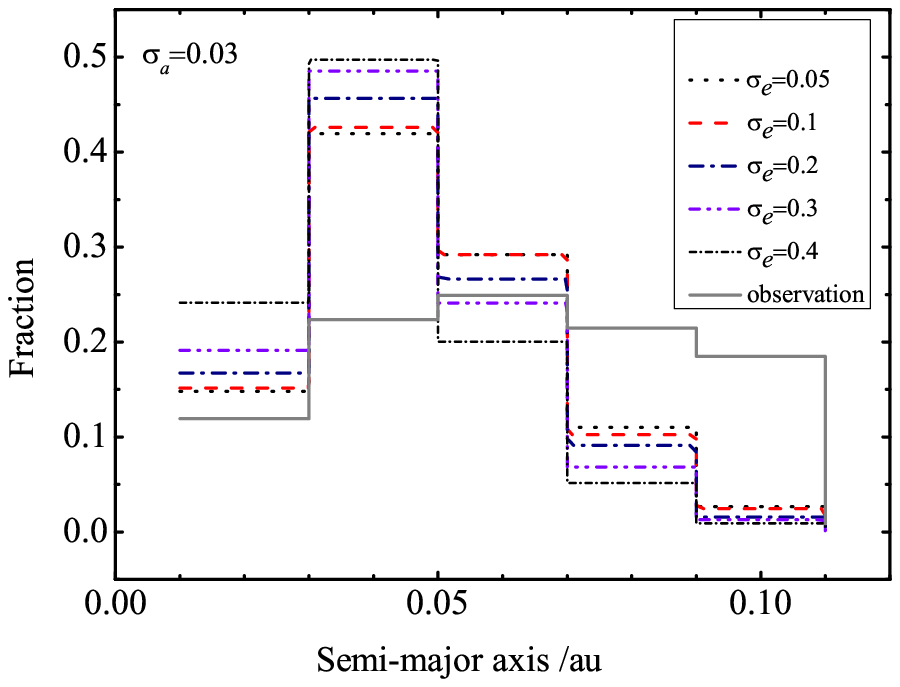}
\caption{The final distributions of semi-major axis corresponding to the case of $\sigma_a$ = 0.03,
  under the initial eccentricity distributions  of $\sigma_e=$0.05, 0.1, 0.2, 0.3 and 0.4, respectively.
  The solid line represents the observed data.}
    \end{figure}

Fig.4 shows the results of the case of $\sigma_a$ $=$ 0.05.
The peaks in the cases of $\sigma_e$ $=$ 0.05, 0.1, 0.2 are distributed near 0.05 au,
which is basically consistent with the observation.
Especially, the two curves of the lower eccentricity cases ($\leq$ 0.1) are overlapped.
However, when $\sigma_e >$ 0.2, the peak position of semi-major axes is moved to 0.03 au,
and the change rate of peak value attains 38$\%$ for the case of $\sigma_e$ $=$ 0.4.
All the peaks have a similar trend that the peak value of semi-major axes increases with the increase of initial eccentricity.
Hence, we infer that most of low-mass \textit{Kepler} planets may form near 0.05 au by the \emph{in situ} formation scenario, rather than the planet-planet scattering.
\begin{figure}[tbph]
\centering
\includegraphics[scale=1.2]{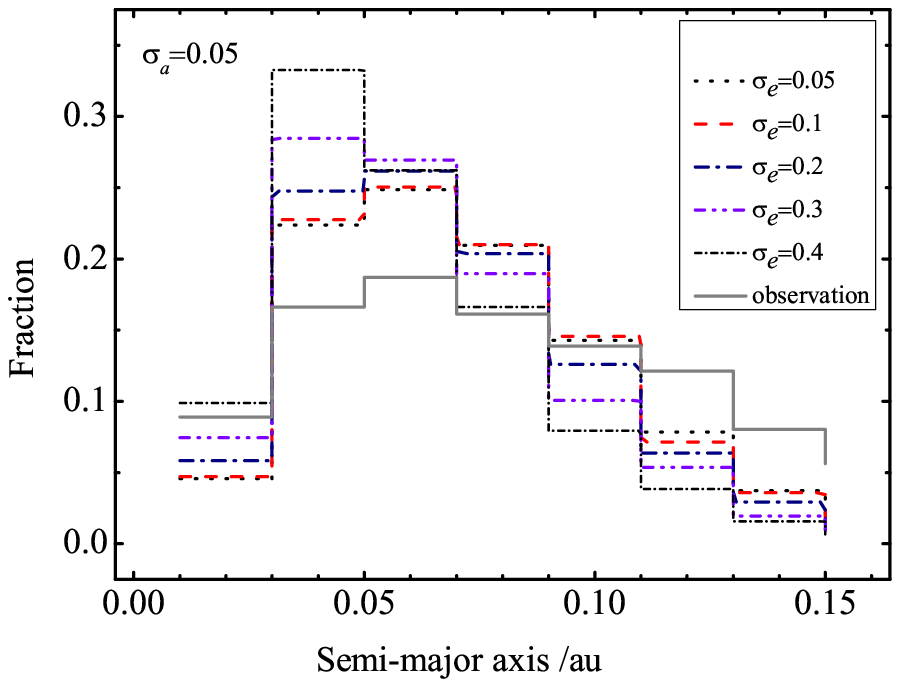}
\caption{Similar to Fig.3, the final distributions of semi-major axis corresponding to the case of $\sigma_a=$ 0.05.}
    \end{figure}

\begin{figure}[tbph]
\centering
\includegraphics[scale=0.5]{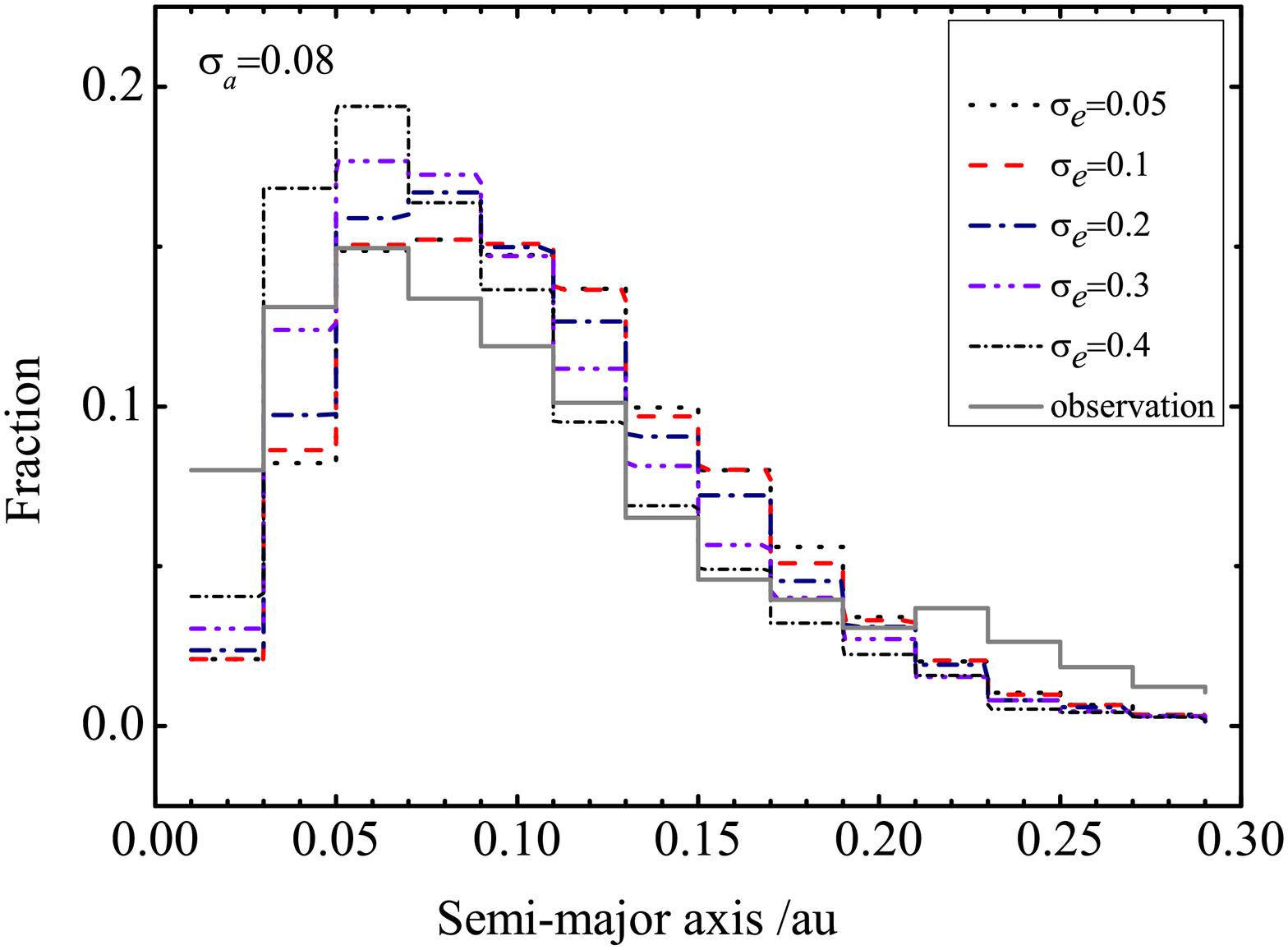}
\caption{Similar to Figs.3-4, the final distributions of semi-major axis corresponding to the case of $\sigma_a=$ 0.08. }
    \end{figure}

For the case of $\sigma_a$ $=$ 0.08, the simulation results are shown in Fig.5.
The two curves with $\sigma_e$ $<$ 0.2 are very similar to the initial one,
which indicates that the tidal interaction is very weak.
When $\sigma_e$ $>$ 0.2, $a$-peaks move to 0.05 au from the initial 0.08 au,
causing more planets piled up near 0.05 au, which is similar to the observations.
This indicates that the result of tidal evolution depends strongly on the initial eccentricity distribution.
We speculate that these Earth-like planets are very possibly piled up near 0.1 au with a low/medium eccentricity before tidal evolutions.

\section{EFFECTS OF VARIOUS PARAMETERS}

\subsection{Evolution Time}

At first, we present the tidal evolutions of planetary orbit in the timescales from million to gillion years.
Fig.6 shows the tidal evolutions of the semi-major axis and eccentricity for the
 evolution timescales of $10^6$, $10^7$, $10^8$ and $10^9$ yr.
The initial orbital semi-major axis and eccentricity are assumed to be randomly distributed between 0.01 au and 0.12 au and between 0 au and 0.4 au, respectively.
About 50$\%$, 76$\%$, 92$\%$, 99$\%$ of planets
have been in circular orbits ($e$ $<$ 0.02) in the above evolution timescales, respectively.
The simulation results show that after a planet experienced the circularization for several million years,
the stellar tide will domain the succeeding dynamical evolution, making the planet increasingly close to the star.

\subsection{Tidal Dissipation Coefficient}

The planetary and stellar tidal dissipation coefficients in the simulations are two uncertain parameters.
As is known that in the tidal evolution, $Q_p^{\prime}$ actually has no influence on the final semi-major axis,
but it plays an important part in the circularization timescale of a planet,
the effect of $Q_p^{\prime}$ on the eccentricity just changes the evolution timescale, but does not affect the amplitude of eccentricity\rf{20,30}.
As shown in Fig.7,
the difference of the semi-major axis distribution
is caused only by the difference of eccentricity distribution, but irrelevant to $Q_p^{\prime}$.
\begin{figure}[tbph]
\centering
\includegraphics[scale=0.5]{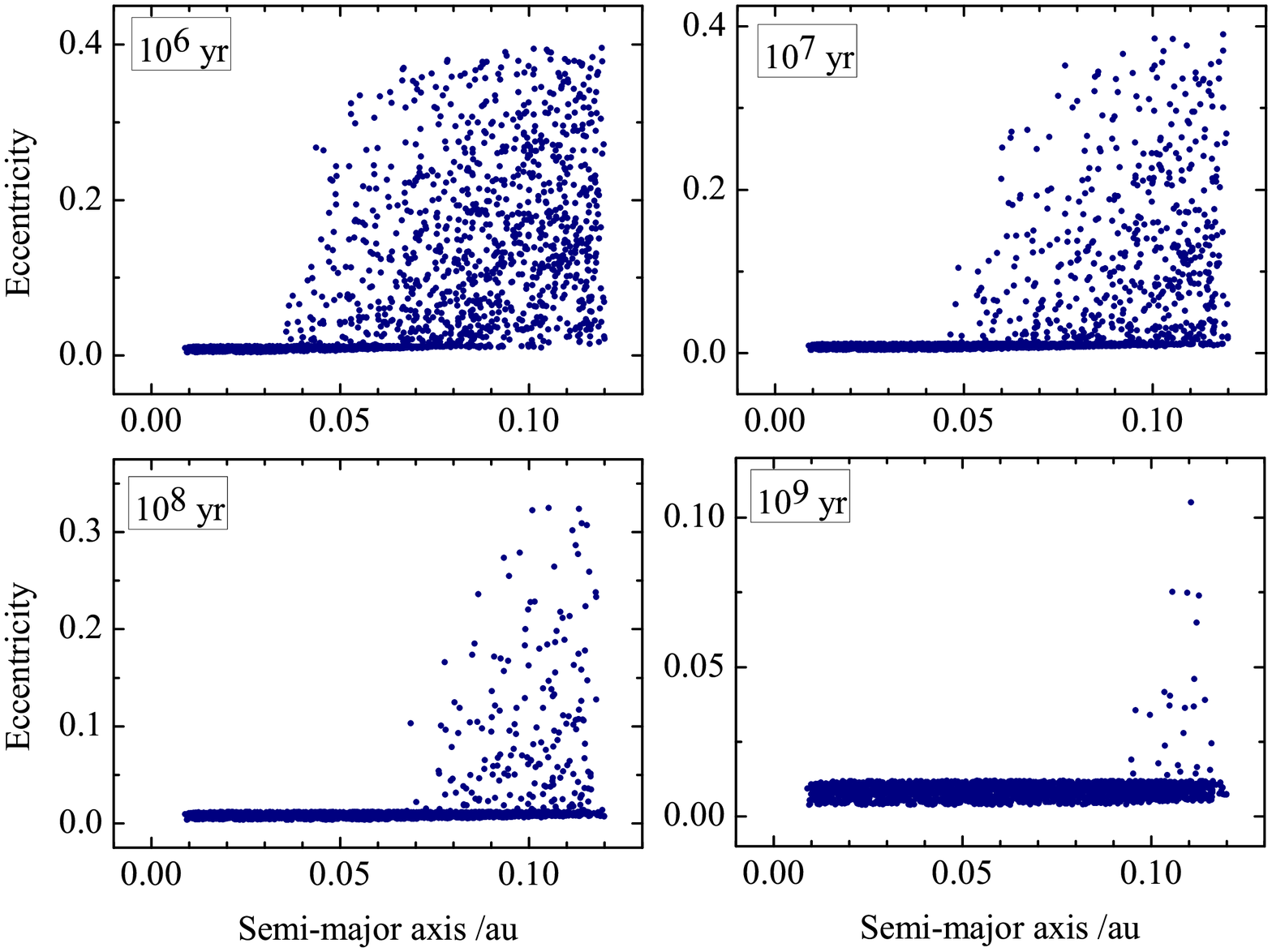}
\caption{The final orbital semi-major axis and eccentricity distributions after undergoing the tidal evolutions of different timescales.
  The initial orbital semi-major axis and eccentricity are chosen randomly from 0.01 to 0.12 au and between 0 and 0.4.
 Planets within 0.1 au are almost circularized after gillion years' evolution.}
    \end{figure}

\begin{figure}[tbph]
\centering
\includegraphics[scale=1.2]{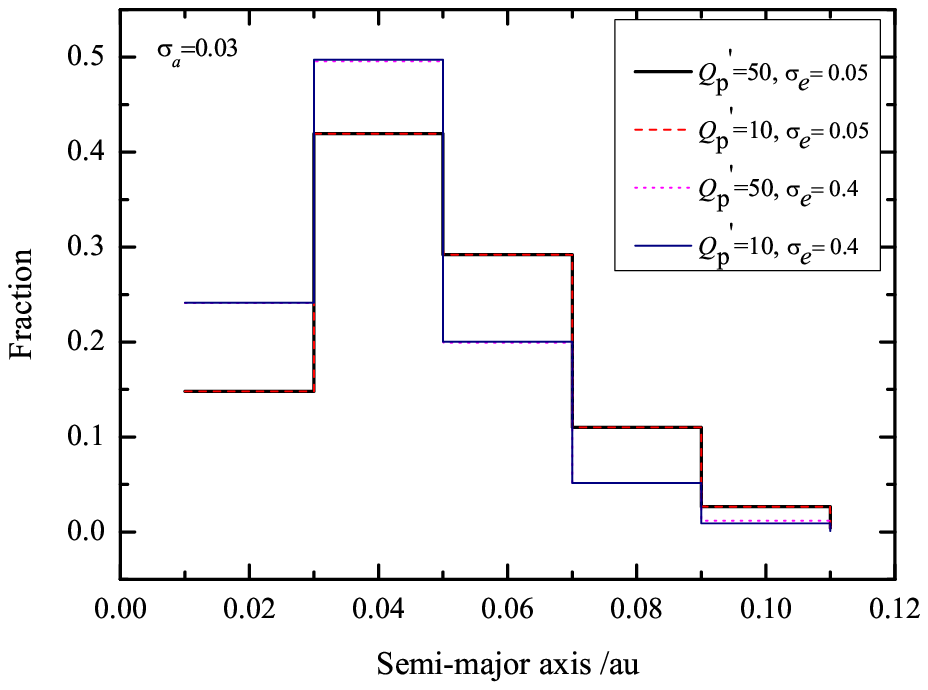}
\caption{The final distributions of semi-major axis with the different values of $Q_p^{\prime}$ and initial eccentricity, which
   depend on the initial eccentricity but not on $Q_p^{\prime}$. }
    \end{figure}

On the contrary, the stellar tidal dissipation coefficient can make a contribution to
the orbital decay, even if the planetary orbit has been circularized.
Therefore, we modified it as $Q_*^{\prime}$ = $10^9$
for the case of $\sigma_a$ $=$ 0.03 to study the effect of $Q_*^{\prime}$.
Fig.8 shows the tidal evolution curves when  $Q_*^{\prime}$ = $10^9$ and $10^7$.
The peak value of semi-major axis decreases with the increase of $Q_*^{\prime}$.
Furthermore, to better understand the effect of $Q_*^{\prime}$,
we compare the final semi-major axis distributions under the different initial eccentricity distributions.
Results show that the initial eccentricity distribution has a greater effect on
the planet's final position than $Q_*^{\prime}$.
The peaks are overlapped at the same position,
but a higher initial eccentricity  and smaller $Q_*^{\prime}$ cause a higher peak value.
\begin{figure}[tbph]
\centering
\includegraphics[scale=0.5]{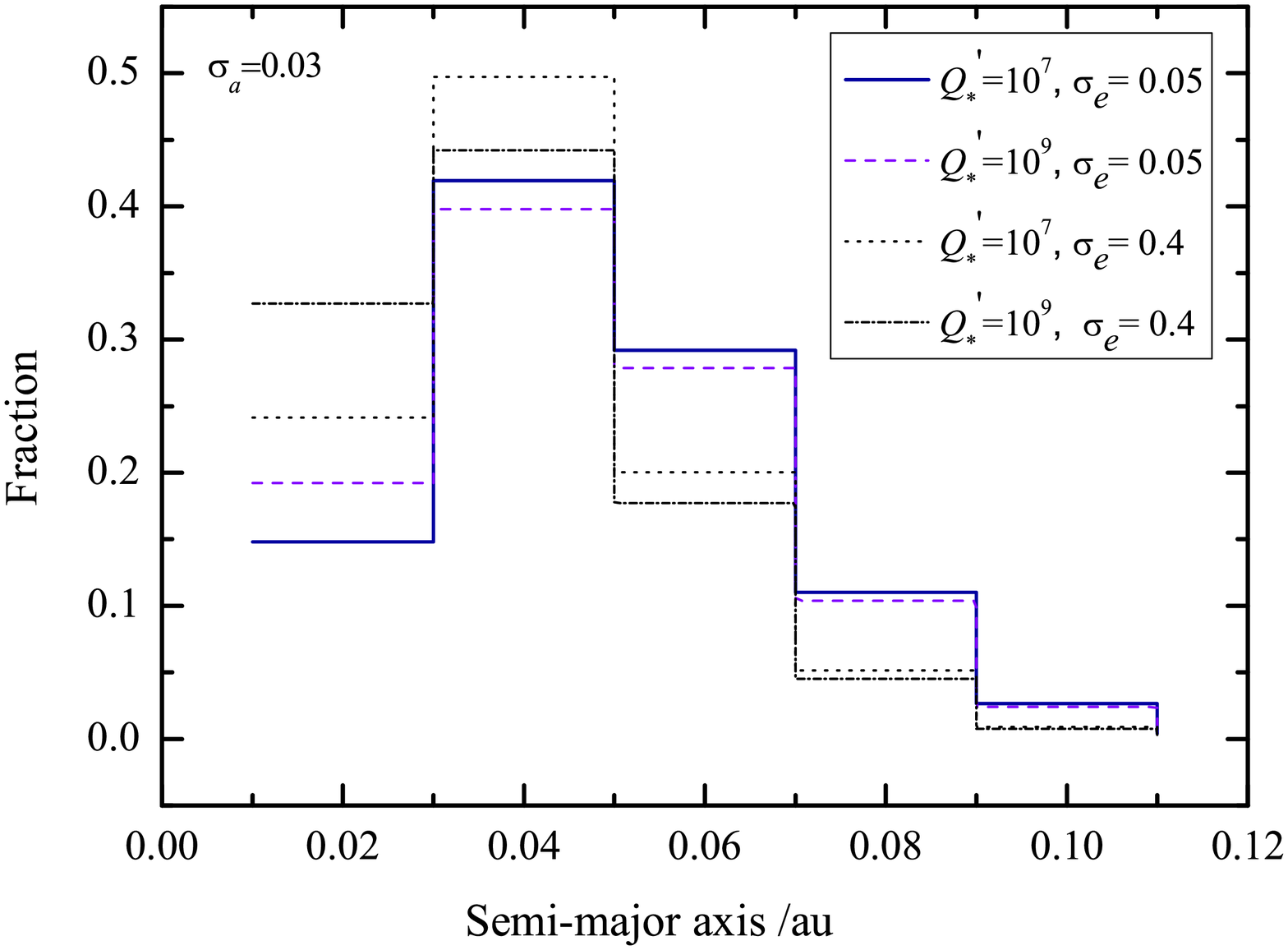}
\caption{The final  distributions of semi-major axis with the different values of $Q_p^{\prime}$ and initial eccentricity.
  The dependence of $Q_*^{\prime}$ is relatively weaker than that of the initial eccentricity distribution.}
    \end{figure}

\begin{figure}[tbph]
\centering
\includegraphics[scale=0.5]{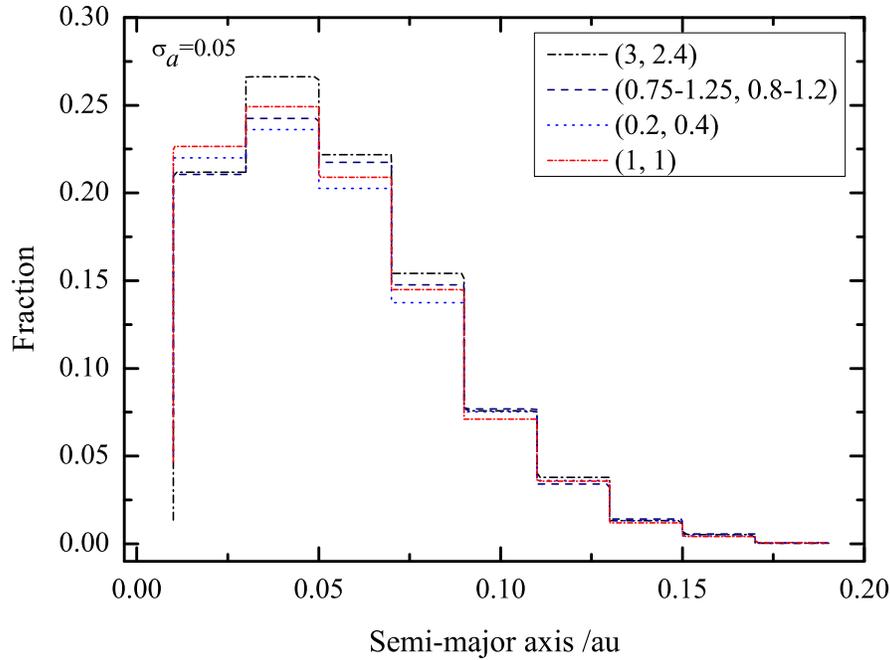}
\caption{The result of tidal evolution for the different stellar mass-radius relations.}
    \end{figure}

\begin{figure}[tbph]
\centering
\includegraphics[scale=0.5]{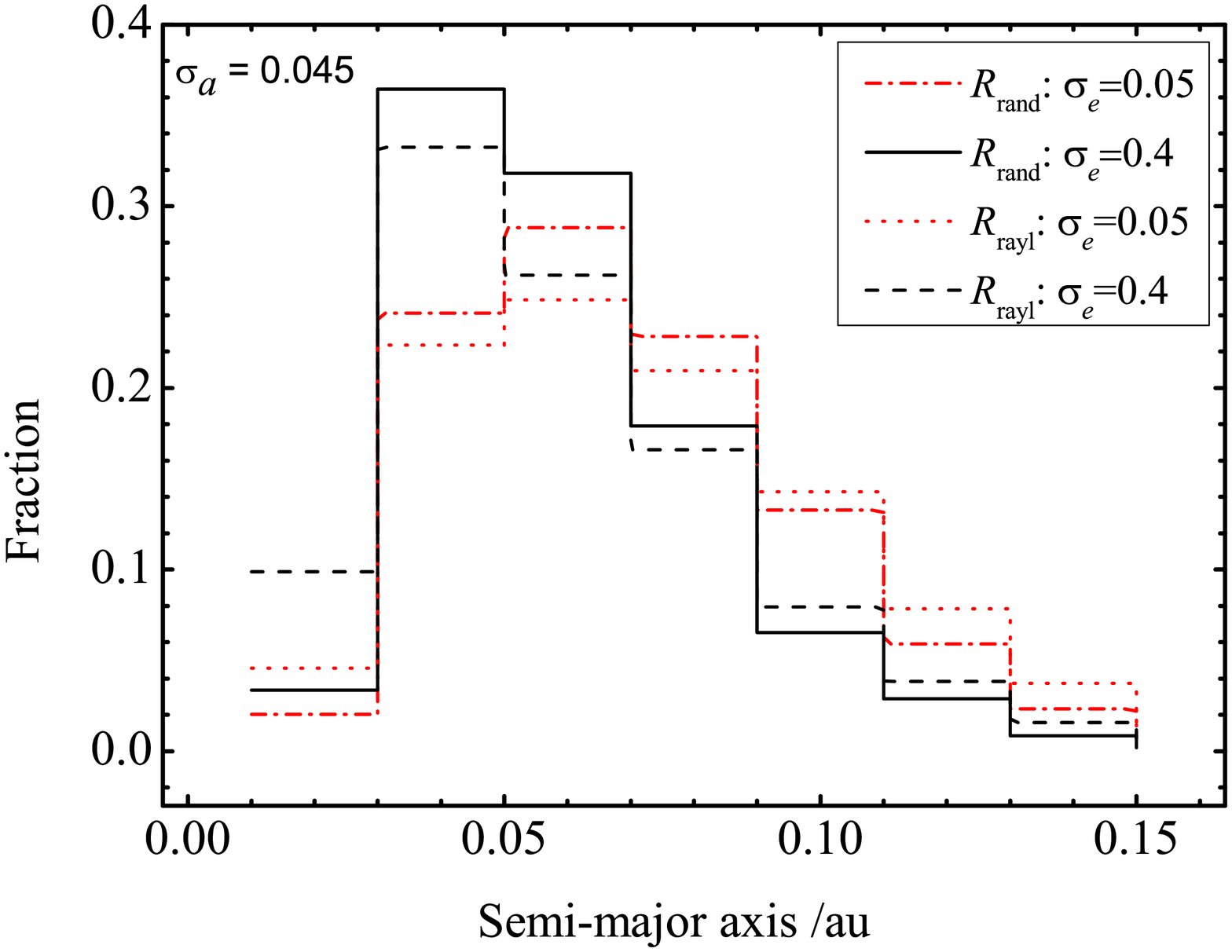}
\caption{The final orbital distribution with different planet radius: $R_{\mathrm{rand}}$ case means $R_p$ is assumed as random values chosen
  from 0.8 to 4 $R_\oplus$, and $R_{\mathrm{rayl}}$ as Rayleigh distribution with $\sigma_R$ =1.3 $R_\oplus$.}
    \end{figure}

\subsection{Mass and Radius}

We investigate first the planetary tidal evolution for the different stellar mass and
radius relations. We adopt the relation of the stellar radius and mass for the main-sequence stars, namely
 $R/R_\odot=(M/M_\odot)^\lambda$, with the parameter $\lambda$ chosen between 0.5 and 1 \rf{31}.
Three $\lambda$-values are selected to obtain the three $(R,M)$ pairs:
(3, 2.4), (0.75-1.25, 0.8-1.2) and (0.2, 0.4), the simulation results are shown in Fig.9.
We find that the stellar mass contributes little to the peak position,
but has a little impact on the magnitude of peak value.
And the evolution result obtained by taking the solar mass and radius as the stellar parameters
is very similar to the calculated result of the second case.

Next, the planet radius is assumed to be randomly distributed in the range from 0.8 to 4 $R_\oplus$, we compare
the obtained simulation result with that of the Rayleigh distribution of $\sigma_R$ = 1.3 $R_\oplus$ as shown in Fig.10.
and they are denoted as $R_{\mathrm{rand}}$ and $R_{\mathrm{rayl}}$, respectively.
The final semi-major axis distributions have the similar peak positions for the same eccentricity distribution.
However, the results differ apparently with the eccentricity distribution under
 the same planetary radius distribution.
Hence, the initial distribution of planetary radius will only change the magnitude of peak value
but has no crucial effect on the peak position.
This indicates that the effect of planetary radius on the final distribution of orbital semi-major axis is not large.

In summary, the tidal evolution of a planet depends on the orbital and physical parameters of the planetary system,
and the orbital semi-major axis and eccentricity are the essential ones.
Because that the tidal evolution is a process of energy dissipation, in the case without the action of other external forces,
the total angular momentum of the system is conservative, hence the tidal force always makes the planet orbit attenuated and
circularized.
The physical parameters of the planets and host stars, as well as
the tidal dissipation coefficients only have a very small effect on the final orbital distribution of planets.
In this work, most of the planetary orbits are circularized after a few gillion years' evolution.
The tidal dissipation coefficient represents the strength of tidal dissipation.
A smaller tidal dissipation coefficient  means a strong tidal dissipation, reversely the tidal dissipation is weak.
Both the tidal theory and simulation result indicate that the tidal dissipation coefficient of a planet has no effect
on the final semi-major axis after the planet has been circularized.

\section{Conclusion and discussion}

We have performed the numerical calculations for the tidal evolutions of a large number of short-period
low-mass exoplanets based on a classical tidal model, and presented the characteristics of planetary orbit distribution after tidal evolution:
the peak of orbital semi-major axis distribution moves to a farther position with the increase of initial semi-major axis;
for the same initial semi-major axis distribution, after undergoing the tidal evolution, the number of planets aggregated at the peaks increases
with the increase of initial eccentricity.

In details, when the initial mean semi-axis is 0.03\,au, all the peaks of semi-major axis are overlapped at the same position,
and for the other values of initial mean semi-major axis, the peaks are distributed between 0.03 au and 0.05 au when $\sigma_e$ is larger than 0.2.
Our simulations show that when $\sigma_a$ = 0.08 and $\sigma_e$ $>$ 0.2,
the final distribution of semi-major axis is relatively consistent with the observation,
but there is still a certain difference in the proportion distribution.
We speculate that two reasons may cause such difference:
some parameters used in our simulations are different from those of observation data;
and the number of low-mass planets should be revised, because a part of low-mass planets on close orbits have not yet been discoverred\rf{8}.
At the same time, we have studied the effects of the tidal dissipation coefficient, the mass and other parameters of the planet and star on
the tidal evolution.
We find that all these parameters only affect slightly the peak value of the orbital semi-major axis distribution,
but will not affect its peak position.

Finally, we have tried to explore some clues for the formation mechanism of these short-period low-mass planets.
We suggest that most planets cannot pile up around 0.05 au via any formation mechanism before tidal evolution,
where the strong tidal interaction could drive the planet
into a closer orbit or tear off its major part.
The planets may aggregate at a farther location (for example, $\sim$ 0.1 au)
with a moderate eccentricity.
The simulation result indicates that by the type I migration in the protoplanetary disk, a low-mass planet can
enter into the region within 0.1 au\rf{32-33},
but it does not mean that all planets form near the snow lines. The high occurrence rate of sub-Neptune-mass planets within 0.25 au seems to support
the \textit{in situ} formation scenario\rf{17}.
However, all the planet formation theories need to be further verified in the future.
We may assume that these planets of medium eccentricity form after the type I migration or form \textit{in situ}
to reach the places near 0.1 au,
then experience some unknown mechanisms, for example,
the gravitational perturbation of companions excites a considerable eccentricity in a multiple-planet system,
and finally under the tidal action they approach to the places within 0.05 au.
Later, we will investigate the tidal evolutions of multiple-planet systems,
where the coupling between the tidal interaction and the gravitational disturbance may affect the final orbital
distribution of planets, and form different orbital configurations.

\end{document}